\documentclass[twocolumn,showpacs,aps,nofootinbib,floatfix]{revtex4}

\usepackage{graphicx}

\usepackage{latexsym}
\usepackage{amssymb}
\usepackage{amsmath} 
\usepackage{bm}
\usepackage[rightcaption]{sidecap}
 
\usepackage{graphicx} 
\graphicspath{ {images/} }
\usepackage{array}

\usepackage[center]{subfigure}

\begin{document}

 \newcommand{\bq}{\begin{equation}}
 \newcommand{\eq}{\end{equation}}
 \newcommand{\bqn}{\begin{eqnarray}}
 \newcommand{\eqn}{\end{eqnarray}}
 \newcommand{\nb}{\nonumber}
 \newcommand{\lb}{\label}
\newcommand{\PR}{Phys. Rev. }

\title{Charge of dust particles in a particle chain}

\author{Razieh Yousefi }
\email{Raziyeh\_Yousefi@baylor.edu}

\author{Mudi Chen}
\email{Mudi\_Chen@baylor.edu}

\author{Lorin S. Matthews}
\email{Lorin\_Matthews@baylor.edu}

\author{Truell W. Hyde}
\email{Truell\_Hyde@baylor.edu}

\affiliation{ CASPER, Physics Department, Baylor
University, Waco, TX 76798-7316, USA\\
}
\date{\today}

\begin{abstract}

Charged dust particles form structures which are extended in the vertical direction in the electrode sheath of a rf discharge when confined within a glass box. The charge on each particle as a function of height varies due to the changing plasma conditions and the wakefield of upstream particles. Here an analysis of the equilibrium state of chains of varying number of particles is analyzed to determine the charge on each particle within a vertically extended chain as well as the magnitude of the positive wakefield charge.

\end{abstract}


\maketitle

\section*{INTRODUCTION}
\renewcommand{\theequation}{1.\arabic{equation}} \setcounter{equation}{0}
Dusty (complex) plasmas have attracted increasing interest in the last several years due to their presence in numerous space environments as well as laboratory devices and industrial processes \cite{1a,1b,1c,1d}. Formation of dust crystals in a laboratory dusty plasma, which was first predicted theoretically \cite{1e} and later found experimentally \cite{1f,1g}, provides the possibility to examine crystalline structures and their dynamics on macroscopic time and spatial scales \cite{1,2,3}. 

Dust particles are charged by collecting electrons and ions from the plasma, generally become negatively charged due to the higher mobility of electrons \citep{4}. In laboratory RF plasmas, these particles are trapped in the sheath of the plasma, above the lower electrode where an inhomogeneous electric field in the vertical direction levitates the particles against gravity \citep{5,6,8}. Due to the strong confinement  provided by the plasma sheath in the vertical direction, ordered structures tend to be two-dimensional, such as a plasma crystal \citep{9,10,10a,11,12} or horizontal clusters \cite{12a}. Placing a glass box on the lower electrode increases the horizontal confinement, allowing structures to be formed which are extended in the vertical direction.  The horizontal confinement can be adjusting by changing the system power and pressure, allowing single or multiple vertical chains to be formed.

The stability of these vertically aligned structures is not easily explained on the basis of a repulsive interparticle potential and appears to be dependent on the balance between an attractive ion wakefield and repulsive screening Coulomb potentials \citep{3,13}. In the region of the sheath-plasma interface
above the lower electrode, positive ions accelerate downward toward the lower electrode. The ion flow, deflected by the negatively charged grains, leads to the formation of excess positive space charge just below the negatively charged dust grains levitated in the sheath \citep{14}. Based on numerical calculations, the positive space charge becomes weaker for the downstream grains since a significant fraction of ions has been deflected by upstream grains so that their direction of motion is no longer along the direction of unperturbed ion flow \cite{15}. In addition, the downstream particles are apparently less charged, resulting in a smaller electrostatic lens \citep{15a}.         

Understanding the dynamics of two or more particles in a vertically extended structure is complicated as it is necessary to have an understanding of charging of the particles in such arrangements. Charging of a single particle in the presence of the ion flow has been studied both theoretically by considering a wake field potential generated by an upstream dust grain, which results in attracting other grains to the stationary points behind the grain \citep{16} and also numerically in a flowing plasma \citep{9,17,18,19,20,21,22}. In previous works the dynamics of two vertically aligned particles have been studied experimentally by analyzing the particles' trajectories \citep{8,23,24} and investigated numerically for two vertically aligned dust particles in a flowing plasma \citep{15,22,24}. In \citep{15}, the charging and wakefield structure of three aligned particles was modeled numerically by adding a third particle to the two particles aligned with the flow. Using the equilibrium distances between the two upstream particles, the forces acting on the third particle due to the upstream grains and ion focusing were analyzed to determine that the interparticle distance between the second and third grains is generally less than that between the upper grains.    

A complex structure consisting of multiple particles, such as a long chain, makes it more difficult to quantify the charge distribution on the dust particles as well as the positive space charges formed as a result of the ion flow. In this study, the dynamics of multiple dust particles in vertically aligned particle chains observed experimentally are analyzed numerically using a model based on the ion flow and the formation of positive space charges below the top particles in the chain. An iterative procedure is used to obtain the best values for charges on the particles as well as the positive space charges and their positions. These results are compared with previous studies.

 \section*{EXPERIMENT} 

\begin{figure}
\includegraphics[width=8cm]{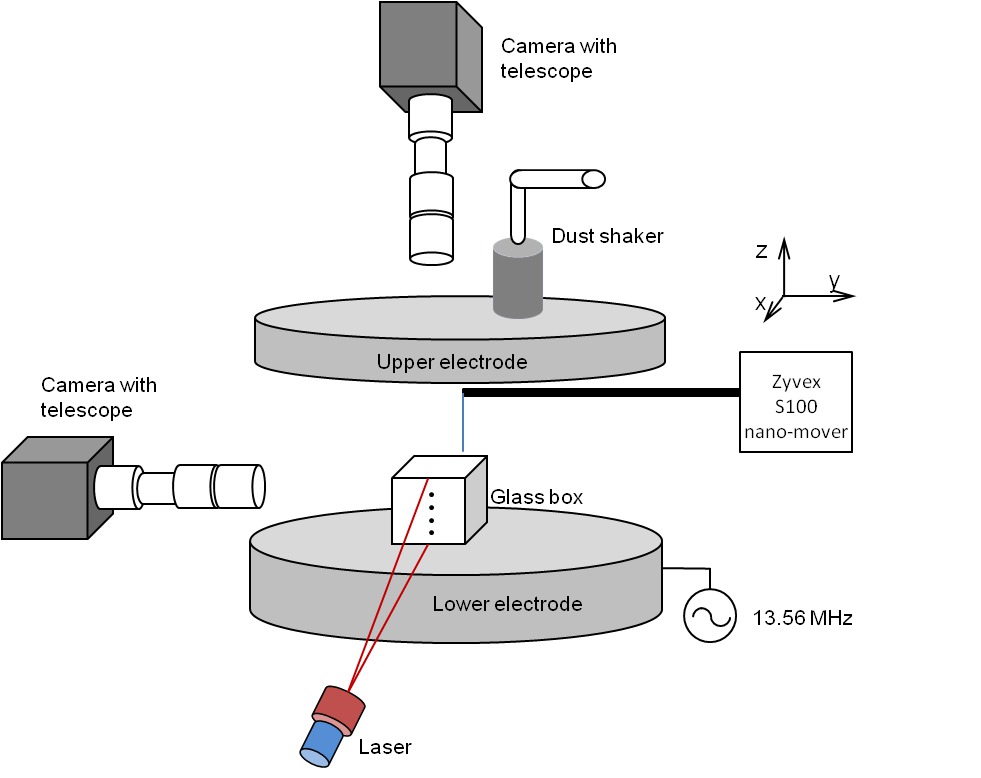}
\centering
\caption{(Color online) A schematic of the experimental setup. The open-ended glass box sitting on the lower electrode has dimensions of 12 mm $\times$ 10.5 mm. Dust particle chains of different lengths were formed inside the box at different operating powers.}
\label{fig1}
\end{figure}

The experiment presented here was conducted in a modified GEC (Gaseous Electronics Conference) RF reference cell. The cell has two electrodes, a lower cylindrical electrode driven at 13.56 MHz and a hollow cylindrical upper electrode, which is grounded. As shown in Fig. \ref{fig1}, an open-ended glass box is placed on top of the lower electrode in order to provide horizontal confinement for the dust particles \citep{11}. The experiment is performed in an argon plasma at a pressure of 100 mTorr with the RF power set at 6.5 W. Melamine Formaldehyde (MF) spheres with diameter of 
8.94$\pm$0.09 $\mu$m and mass of 6.10$\pm$0.09 $\times 10^{-13}$ kg were introduced into the argon plasma, using shakers mounted above the hollow upper electrode, and imaged at 500 frames per second  using a side-mounted, high-speed CCD (Photron) camera and a microscope lens.

A single vertical dust particle chain is formed at the center of the glass box with the RF power set at 6.5 W. By slowly decreasing the RF power, the lowest particle in the chain is removed, leaving a chain consisting of six particles. The RF voltage is then returned to its original magnitude and the particle positions are recorded. By lowering the power again, the length of the chain is shortened by removing the lowest particle from the chain \cite{26}. Restoring the power to its original setting allows a vertical chain consisting of five, four, three, two and one particles to be formed respectively, as shown in Fig. \ref{fig2}.

\section*{ANALYSIS}

In order to study the charging of different particle arrangements in a flowing plasma, a model is constructed based on the formation of a positive space charge below the particles in a chain as a result of the ion flow. The positive space charge becomes weaker for the downstream grains for two reasons: first, the downstream particles become less charged, resulting in a smaller electrostatic lens \cite{15}, and second, the velocity of the ions increases as they approach the lower electrode, resulting in larger momentum, which makes it harder for negatively charged dust particles to affect the ion trajectory to form a positive space charge. Therefore, a positive space charge is assumed to exist only beneath the two top particles in a chain. Fig. \ref{fig3} shows a schematic of the model which is used to study the charging of the particles in a vertical chain.

In the experiment, gravity pulls the grains toward the lower electrode, while within the sheath a strong vertical electric field arising from the negatively charged lower electrode pushes them upward. The interaction force between negatively charged dust grains is a shielded Coulomb interaction, as has been considered in many previous calculations \cite{27a,27,28}. For simplicity, the electric force arising from the positive space charges is assumed to be a Coulomb interaction which attracts the downstream negatively charged dust particles. 

\begin{figure}
\includegraphics[width=.5\textwidth]{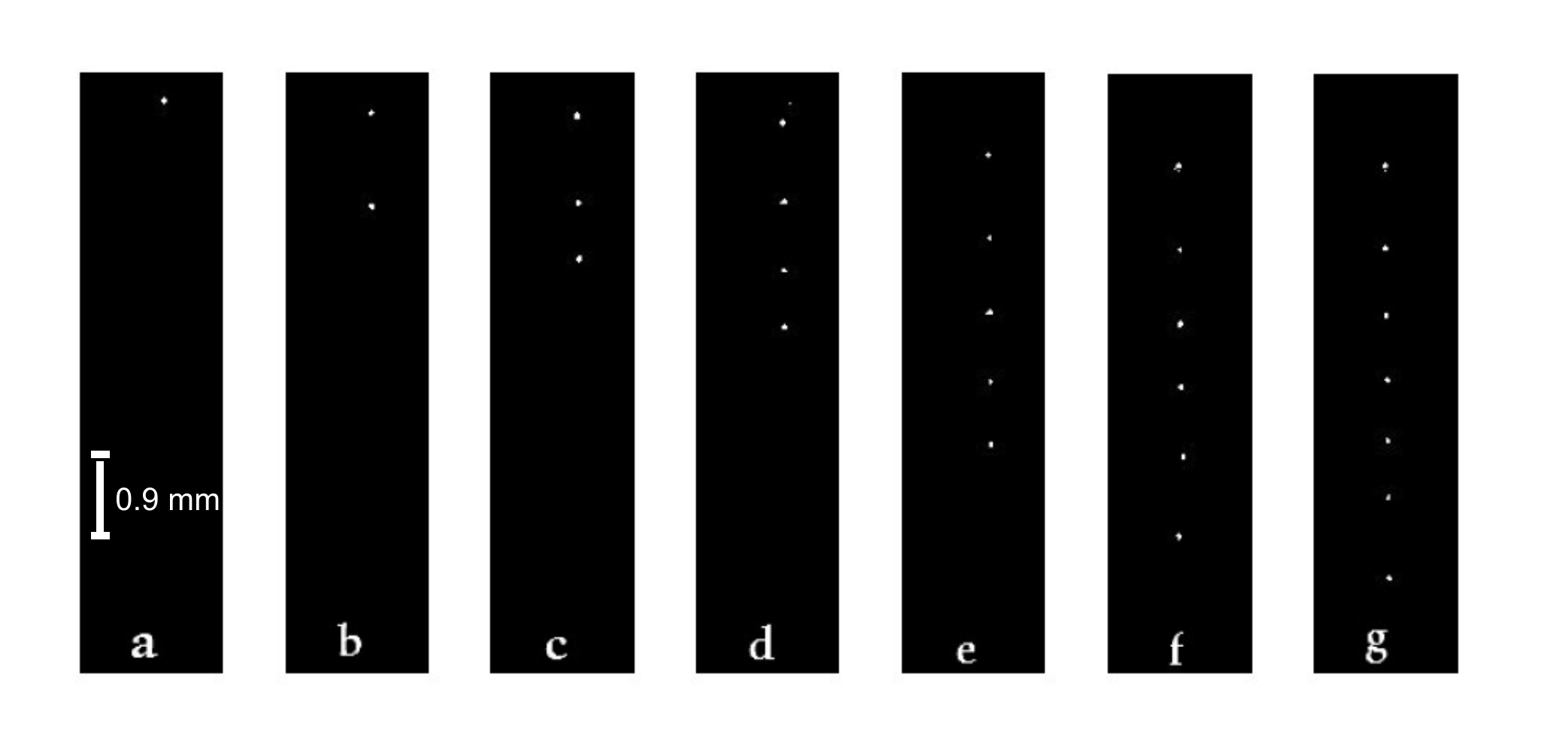}
\centering
\caption{The vertical alignment of single-particle to seven-particle chains, left to right. Vertical position of dust particles are measured from this data.}
\label{fig2}
\end{figure}

Due to the experimental geometry and the vertical structure of a chain, only a 1D set of equations is needed to describe the interparticle forces and analyze the charges present in the system. For the $i^{th}$ particle in a chain consisting of $n$ particles, the total force, $\vec{F}_{tot_i}$, is

\begin{equation}
\lb{1}
\vec{F}_{tot_i} = \vec{F}_{G_i}  + \vec{F}_{z_i} + \vec{F}_{int_i} 
\tag{1}
\end{equation}
where $\vec{F}_{G_i} = -m_ig\hat{z} = -mg\hat{z}$ is the gravitational force on the $i^{th}$ dust particle with mass $m_i=m$, $\vec{F}_{z_i}$ is the electrostatic force arising from the lower electrode and $\vec{F}_{int_i}$ is the electrostatic interaction force between the particles. 

It is assumed that the vertical electric field in the vicinity of dust particles is linear in the first order, $E(z) = \alpha z$, as determined for the same experimental setup \cite{31} and thus the vertical electric force acting on each dust particle is given by

\begin{equation}
\lb{2}
\vec{F}_{z_i} = q_i \alpha z_i 
\tag{2}
\end{equation}
where $q_i$ is the charge on the $i^{th}$ particle at vertical position $z_i$ in a coordinate system with the origin at the top of the box, and $\alpha$ is a constant variable to be derived. 

The particle-particle interaction force, $\vec{F}_{int_i}$, is the electrostatic force on the $i^{th}$ particle in the chain arising from the rest of the charged dust particles as well as the assumed positive space charges, as given by the following equation

\begin{equation}
\begin{split}
\label{3}
\vec{F}_{int_i} & = -\frac{q_i}{4\pi\epsilon_0} \Sigma_{j=1}^{n-1} [ \frac{q_j}{z_{ij}}(\frac{1}{z_{ij}}+\frac{1}{\lambda}) \exp(\frac{-z_{ij}}{{\lambda}})\frac{\vec{z_{ij}}}{z_{ij}} ]  \\ 
& +\frac{q_i}{4\pi\epsilon_0} \Sigma_{k=1}^{2}  Q_k\frac{\vec{z_{ik}}}{{z_{ik}}^3}
\end{split}
\tag{3} 
\end{equation}

where $z_{ij}=\mid z_i-z_j\mid$, $\vec{z_{ij}} = (\vec{z_i}-\vec{z_j})$, $\lambda_i$ is the screening length (initially assumed to be constant for all the dust particles, and allowed to vary in subsequent iterations), and $n$ is the number of dust particles. In the first part of the equation, the interaction force arising from the $j^{th}$ dust particle is given while the second part represents the electrostatic force due to the assumed positive space charge $Q_k$ located a distance $d_k$ beneath the upper two grains. As the space charges result from the focusing of the ions in the plasma, the additional shielding due to other ions in the plasma is not considered  in calculating this interaction force.
   
The particle chains observed in this experiment are assumed to be stable structures with $\vec{F}_{tot}=0$. The vertical position of each particle, $z_i$, is obtained from experiment as shown in Fig. \ref{fig2}, and these values are substituted into Eqs. \ref{2} and \ref{3}. A numerical method is used to solve the set of equations by setting maximum and minimum initial values for each variable. The actual value of each variable is assumed to lie within this range, and each variable is adjusted until all of the equations are satisfied within an allowed tolerance of error, $\epsilon$,

\begin{equation*}
\lb{4}
\vert F_{tot_i} \vert < \epsilon  ~~~~~~~~~~~~~~~~~(i= 1 ~to~7).
\tag{4}
\end{equation*}

\begin{figure}
\includegraphics[width=.4\textwidth]{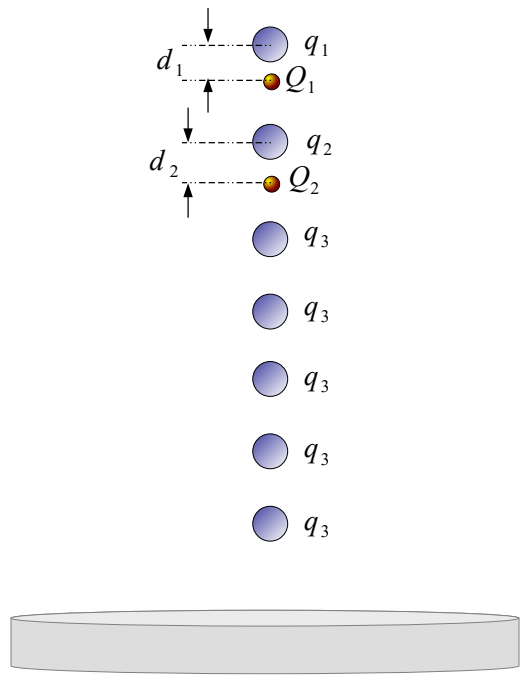}
\centering
\caption{(Color online) A schematic of the model used for studying charging of the dust particles in vertical structures with ion flow. The positive space charges are formed below the two upstream particles. The top particle carries a charge $q_1$ with a positive space charge $Q_1$ below it at distance $d_1$. The second dust particle carries the charge $q_2$ with a positive space charge, $Q_2$, at distance $d_2$ below it. The remaining particles carry the charge $q_3$. The dust particle chain is formed in the sheath above the lower electrode.}
\label{fig3}
\end{figure}

~~~~~~~~~~~~~~~~~~~~~\
~~~~~~~~~~~~~~~
~~~~~~~~~~~~~~~~~~

\section*{RESULTS}
Solving for the charge on $n$ particles plus two positive space charges, with $d_i$, $\lambda$ and $\alpha$ as unknowns requires a set of $n+6$ equations. Since only one equation of the form of Eq. \ref{1} can be obtained for each particle, some simplifying assumptions must be made.

As a first step, it is assumed that all the particles below the top two particles have the same charge, $q_3$. This brings the number of unknowns to nine, and therefore a chain of seven particles in addition to two space charges provides the necessary number of equations. The initial estimate of the unknown values  calculated in this way are provided in Table \ref{table:1} for $\epsilon = \frac{m g}{50}$.

\begin{table}[ht]
\def\arraystretch{1.25}
\caption{Calculated values for a seven particle chain.} 
\centering 
\begin{tabular}{c  c c} 
\hline\hline 
Quantity & Initial Estimate & Final Value\\
\hline
Charges ($10^{4}$ e)  \\
$q_1$ & -3.46  & -3.88\\ 
$q_2$ & -2.73 & -3.20\\
$q_3$ & -2.60 & -2.96\\
$Q_1$ & +0.63  & +0.65\\
$Q_2$ & +0.29  & +0.18\\ 
\hline 
Distances ($\mu$m)  \\
$d_1$ & 180  & 186\\
$d_2$ & 300 & 167 \\
$\lambda$ & 705  & 470-649\\
\hline 
Electric field constant($V/m^2)$ \\
$\alpha$ & -3.07 $\times 10^{-4}$   & -3.08$\times 10^{-4}$\\ 
\hline \hline 
\end{tabular}
\label{table:1} 
\end{table}



From Table \ref{table:1}, the estimated charges on the top two particles in the chain are related by $q_2~ = ~0.79 ~q_1$, which is consistent with previous experimental observation where the downstream particle is measured to have a charge of $78\% $ of the upstream particle, i.e. 
$q_2/q_1 = 0.78$ \cite{30a}. The estimated positive space charge $Q_1$ is found to be related to the charge on the top dust particle by $Q_1 = 0.18~ q_1$ which is in agreement with the numerically predicted value for a positive space charge of $15\%$ of the charge on the upstream particle \cite{15}. In addition, the location of the upper space charge, $d_1$, is 18 $\%$ of the inter-particle spacing $\delta z$, which agrees with the previously reported results for the position of the positive space below the top particle \cite{22,24}. 
Note that $Q_2~ =~ 0.08 q_1$ also provides support for the assumption that smaller positive space charges exist for down-stream particles, as found in previous studies \cite{15}.
The screening length is found to be $\lambda \approx 0.7 \delta z$. This value is in agreement with the calculations for stable aligned structures using a dynamically screened Coulomb interaction between dust grains in a chain \cite{30}. 

The estimated set of solutions is derived for the case where the third through seventh down-stream particles all carry the same charge. In the next step, using the same algorithm, the charges on the down-stream particles are allowed to differ by defining two initial values for each grain based on the charges calculated in the first step. This allows the magnitude of error to become as small as $\epsilon = \frac{mg}{10^4}$ with the final calculated values, which are shown for comparison in Table \ref{table:1}.

\begin{figure}[t]
\includegraphics[width=.45\textwidth]{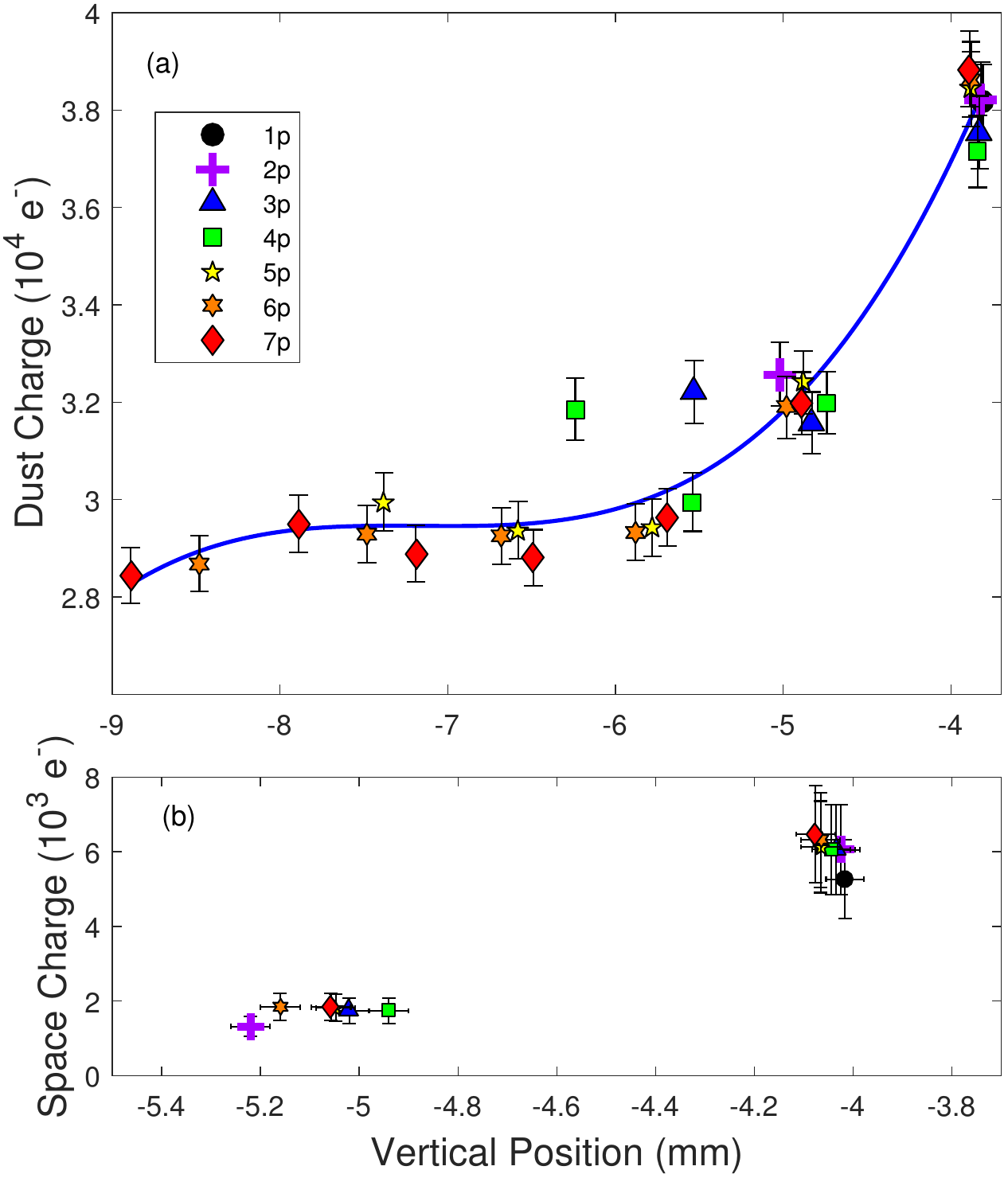}
\centering
\caption{(Color online) a) Charge on the dust particles in vertical chains consisting of one to seven particles, with position measured relative to the top of the box.  The number of particles in the chain is indicated in the legend. The line serves to guide the eye. b) Magnitude and position of the positive space charges below the top two particles in vertical chains, with the length of the chain indicated as in (a). Note the expanded axis for the vertical position.  Error bars denote the upper and lower limits for the derived quantities.}
\label{fig4}
\end{figure}

Given the solution for a seven-particle chain, the numerical method is then used to solve the set of equations for successively shorter chains, where again the ion flow is assumed to create positive space charges only below the the top two particles. In the calculations, none of the parameters are set to be constant, but are allowed to vary within a narrow range about the best-fit values calculated for the previous iteration. The results for the magnitude of the charges on each dust particle in chains of different lengths are provided in Fig. \ref{fig4}a and the magnitude and positions of the positive space charges formed below the top two particles in each chain are shown in Fig. \ref{fig4}b.

Although there is no way to prove that these are the unique solutions to the systems of equations, these results are compatible with previous numerical and experimental studies. For instance, for the two-particle chain, the charge on the down-stream particle is found to be about $85\%$ of the charge on the top particle, comparable to the results derived in \cite{24} by measuring the attractive force between two aligned dust particles. Moreover, the top positive space charge is found to be $16\%$ of the charge on the top particle with its position $17\%$ of the spacing between the two dust particles, results which are in good agreement with previous studies where the stability of aligned structures has been studied by calculating the forces on the dust particles in a flowing plasma \cite{22}. In addition, for the three particle chain, the middle particle has the smallest charge of the three particles in the chain, about $84\%$ of the top particle's charge, while the lowest particle charge is $86\%$ of that of the top particle. A similar relation between the charges for a three particle chain is predicted in \cite{15} using a numerical method in a flowing plasma.

In general, the charge on the dust particles in a chain decreases from top to bottom as expected, since the number density of electrons decreases relative to that for ions in the plasma sheath as the boundary (lower electrode) is approached. As shown in Fig. \ref{fig4}a, in the longer chains ($n = 4-7$), the charge differences between the lower particles is smaller than that for the top particles, indicating a uniform plasma condition in a small sheath region between $4\leq z \leq 7 mm $. This seems to be a unique condition created by the presence of the glass box \cite{25}. Also notable is the fact that in the three to five particle chains, the charge of the lowest particle is greater than that of the particle just above it. The reason for this could be the shadowing effect \cite{29asb1,29asb2}, since the lower particle is just shielded in one direction by the upper particles while the particle above it is shielded from two directions. 

\begin{figure}[t]
\includegraphics[width=.45\textwidth]{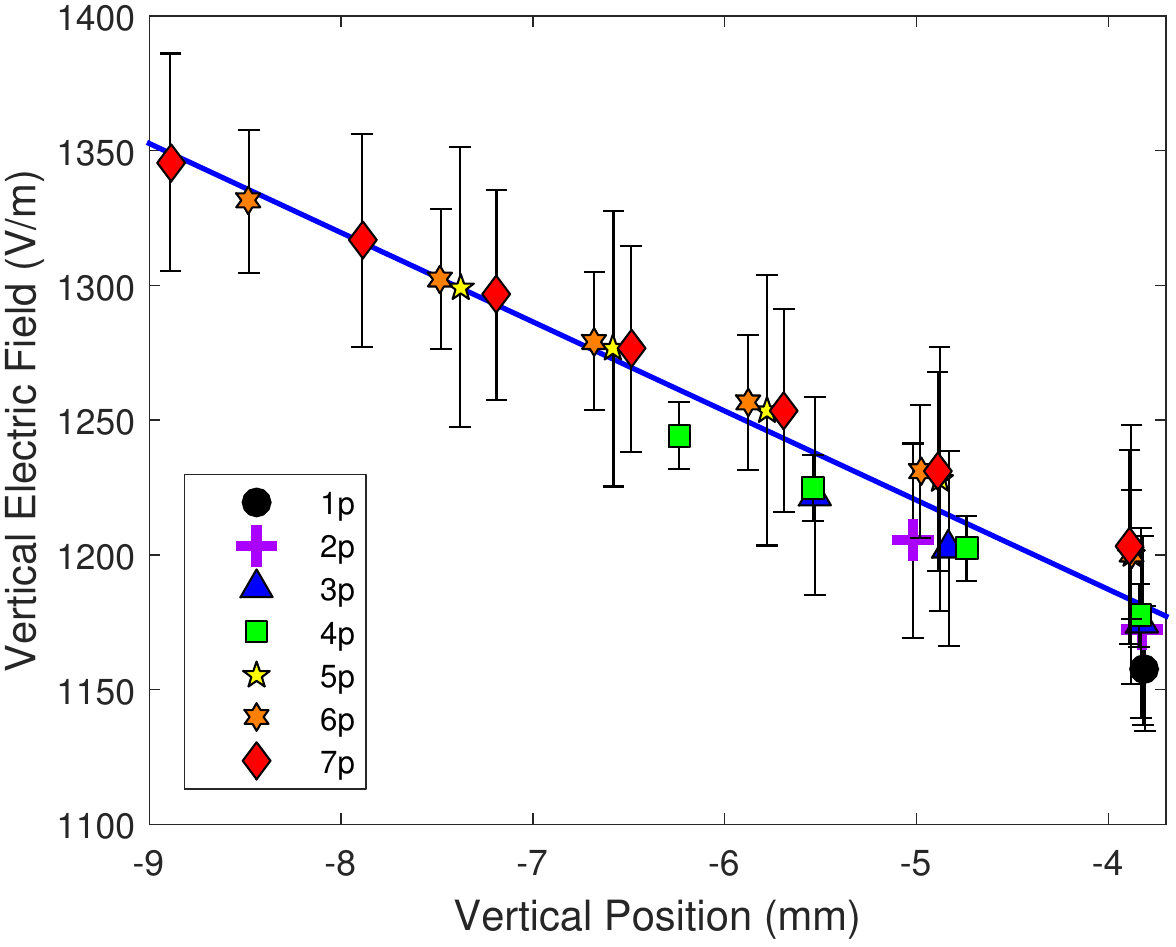}
\centering
\caption{(Color online) Vertical electric field determined at different particle positions for several chains with the same plasma conditions. The electric field is found to be linear in this region of the sheath. The number of particles in the chain is indicated in the legend. Error bars are calculated from the electric field relation considering the lower and upper values for the derived values of $\alpha$.}
\label{fig5}
\end{figure}

The vertical electric field at the position of each particle in the chain is also calculated. As all of the chains are observed at the same plasma conditions, one  expects to find similar electric fields at the same heights. As shown in Fig. \ref{fig5}, for different particle chains the electric field is found to be linear and with similar magnitude for the same positions within each chain, providing a check on the self-consistency of both the numerical method and derived solutions in this work. Also, the magnitude of vertical electric field is of the same order of that derived in the experiment performed in the same RF cell with different plasma conditions \cite{27a} and numerical calculations for the sheath which were made using a fluid model to calculate the bias on the lower electrode and the plasma potential profiles in the sheath \cite{31}.  

\begin{figure}[t]
\includegraphics[width=.45\textwidth]{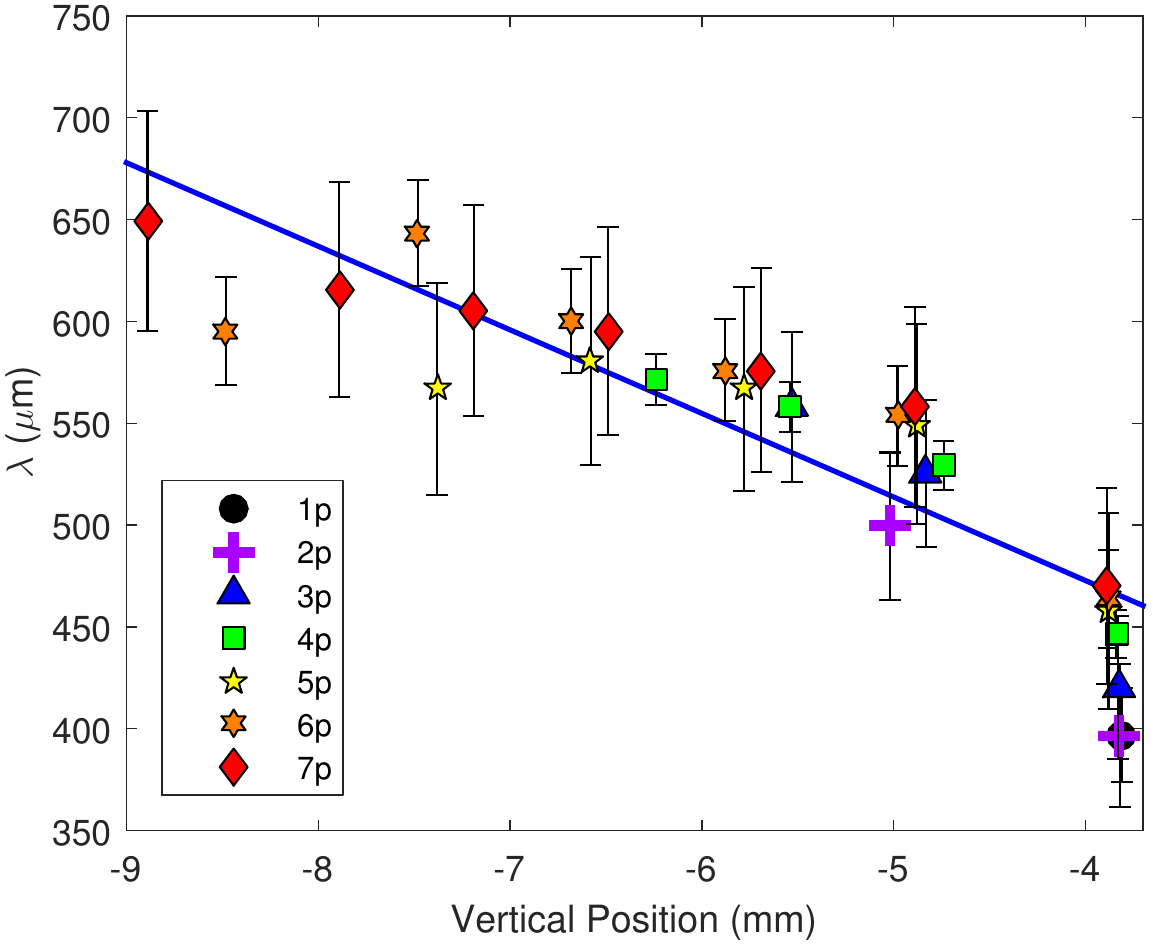}
\centering
\caption{(Color online)Plasma shielding length determined at different particle positions for several chains with the same plasma conditions. The number of particles in the chain is indicated in the legend. Not all of the data points are visible, as some overlap. The solid line is a linear fit to the data.  Error bars denote the upper and lower limits for the calculated values.}
\label{fig6}
\end{figure}

Finally, the plasma shielding length was also varied for each particle within the chain, allowing the changing plasma conditions to be mapped throughout this region of the sheath. Although the shielding length varies continuously with height, a constant shielding length $\lambda_i$ was used in the force calculation for the $i^{th}$ particle. As the dominant contributions to the force are from a particle's nearest neighbors, each value represents the average over the vertical span encompassing adjacent particles. As shown in Fig. \ref{fig6}, the shielding length increases as the particles approach the lower electrode, as the plasma density decreases in this region.  The shielding length calculated for the top particle in each chain may be lower than the linear trend because of the ions focused beneath it.

\section*{CONCLUSIONS}

This study quantified the forces acting on particles in vertically extended chains. Vertical particle chains composed of seven and fewer particles are formed in the sheath of an RF argon discharge plasma, confined within a glass box placed on the lower electrode. The flowing ions are assumed to create positive space charges under the two upstream particles. Treating a chain as a stable system with each particle at rest, the force balance is derived by considering all the forces acting on the particle. In the first iteration, the equations are solved for a seven-particle chain by reducing the number of unknowns, making the assumption that the charge on the third through seventh particles in the chain is the same, and assuming a fixed plasma shielding length. Using the initial estimates obtained in this iteration, all of the parameters in the system of force equations are allowed to vary about these values to minimize the error. Subsequent iterations of the same numerical method, allowing differing values of charges on the particles in each chain, allow the systems of equations to be solved separately for each chain. This allows not only determination of the charges on each particle, put also allows the local plasma parameters to be probed, including the magnitude and position of the positive space charge beneath the top upstream particles, the plasma shielding length for each particle, and the magnitude of vertical electric field at the position of each dust particle. 

The results for short chains ($n \leq 3$ particle chains), are shown to be in agreement with previous numerical and experimental works.  For two and three particle chains, the relationship between the value of charges on the dust particles (as shown in Fig. \ref{fig4}a) as well as the positive space charges and their positions (Fig. \ref{fig4}b) are in good agreement with previous studies \citep{16, 22, 23, 24, 29}.  In particular, for a two particle chain, the value of the charge on the lower particle is $85\%$ of that on the top particle with a smaller posive space charge beneath the lower particle, consistent with the results reported in \citep{24}.  It was also found that the the middle particle in a three-particle chain carries the smallest charge among the three particles, as found in \citep{15}.  The trend that the lowest particle in the chain has a greater charger than the immediate upstream particle continued for the four- and five-particle chains, though this was not found to be the case for the longer chains.  The vertical electric field was determined to vary linearly within the glass box, fig \ref{fig5}, as is commonly found in the sheath above the lower electrode.  In addition, this method allowed the plasma shielding length to be mapped in the vertical direction, as shown in Fig. \ref{fig6}.  These measurements illustrate the use of micron-sized particles as probes to measure the local plasma conditions within the sheath above the lower electrode.

\section*{ACKNOWLEDGMENTS}

This work is supported by the National Science Foundation under Grant No. 1414523.



\begin{thebibliography}{nbound}

\bibitem{1a} M. Horanyi, H. L. F. Houpis, and D. A. Mendis, Astrophys. Space Sci. 144,215 (1988).
\bibitem{1b} C. K. Goertz, Rev. Geophys. 27, 271(1989).
\bibitem{1c} J. Goree, Phys. Rev. Lett. 69, 277 (1992).
\bibitem{1d} M. S. Barnes, J. H. Keller, J. C. Forster, J. A. O' Neill, and D. K. Coultas, Phys. Rev. Lett. 68, 313 (1992).
\bibitem{1e} H. Ikezi, Phys. Fluids 29, 1764 (1986).
\bibitem{1f} J. H. Chu, J. B. Du, and I. Lin, J. Phys. D 27, 296 (1994).
\bibitem{1g} J. H. Chu and I. Lin, Phys. Rev. Lett. 72, 4009 (1994).
\bibitem{1} H. Thomas et al., Phys. Rev. Lett. 73, 652 (1994).
\bibitem{2} V. E. Fortov, V. I. Molotkov, A. P. Nefedov, and O. F. Petrov, Phys. Plasmas 6, 1759 (1999). 
\bibitem{3} K. Takahashi, T. Oishi, K. Shimomai, Y. Hayashi, and S. Nishino, Phys. Rev. E 58, 7805 (1998).
\bibitem{4} S. A. Khrapak et al., Phys. Rev. E 72, 016406 (2005). 
\bibitem{5} J. H. Chu and Lin I, Phys. Rev. Lett. 72, 4009 (1994).
\bibitem{6} A. Melzer, T. Trottenberg, and A. Piel, Phys. Lett. A 191, 301 (1994).
\bibitem{8} A. Melzer, V. A. Schweigert, and A. Piel1, Phys. Rev. Lett. 83, 3194 (1999). 
\bibitem{9} V. M. Bedanov, and F. M. Peeters, Phys. Rev. B 49, 2667 (1994). 
\bibitem{10} A. Melzer, V. A. Schweigert, I. V. Schweigert, A. Homann, S. Peters, and A. Piel, Phys. Rev. E 54, 46 (1996).
\bibitem{10a} L. Candido, J. Rino, N. Studart, and F. M. Peeters, J. Phys. Condens. Matter 10, 11627 (1998). 
\bibitem{11} J. Kong, T. W. Hyde, L. S. Matthews, K. Qiao, Z. Zhang, and A. Douglass, Phys. Rev. E 84, 016411 (2011).
\bibitem{12} M. Kroll, J. Schablinski, D. Block, and A. Piel, Phys. Plasmas 17, 013702 (2010).
\bibitem{12a} T. W. Hyde, J. Kong, and L. S. Matthews, Phys. Rev. E 87, 053106 (2013).
\bibitem{13} V. Steinberg et al., Phys. Rev. Lett. 86, 4540 (2001). 
\bibitem{14} V. Vyas, and M. J. Kushner, J. Appl. Phys. 97, 043303 ( 2005).
\bibitem{15} W. J. Miloch, and D. Block, Phys. Plasmas 19, 123703 (2012).
\bibitem{15a} W. J. Miloch, M. Kroll, and D. Block, Phys. Plasmas 17, 103703 (2010).
\bibitem{16} M. Lampe, G. Joyce, G. Ganguli, and V. Gavrishchaka, Phys. Plasmas 7, 3851 (2000). 
\bibitem{17} E. C. Whipple, Rep. Prog. Phys. 44, 1197 (1981).
\bibitem{18} A. V. Ivlev, G. Morfill, and V. E. Fortov, Phys. Plasmas 6, 1415 (1999).
\bibitem{19} J. Allen, Phys. Scr. 45, 497 (1992).
\bibitem{20} G. Lapenta, Phys. Plasmas 6, 1442 (1999).
\bibitem{21} I. H. Hutchinson, Plasma Phys. Controlled Fusion 45, 1477 (2003). 
\bibitem{22} V. A. Schweigert et al., Phys. Rev. E 54, 4155 (1996).
\bibitem{23} G. A. Hebner, M. E. Riley, and B. M. Marder, Phys. Rev. E 68, 016403 (2003).
\bibitem{24} A. Melzer, V. A. Schweigert, and A. Piel, Phys. Scr. 61, 494 (2000).
\bibitem{25} M. Chen, M. Dropmann, B. Zhang, L. Matthews, and T. Hyde, submitted to Phys. Rev. E (2016).
\bibitem{26} D. He, B. Ai, H. K. Chan, and B. Hu, Phys. Rev. E 81, 041131 (2010).
\bibitem{27a} R. Yousefi, A. B. Davis, J. C. Reyes, L. S. Matthews, and T. W. Hyde, Phys. Rev. E 90, 033101 (2014).
\bibitem{27} G. A. Hebner et al., Phys. Rev. Lett. 87, 2350011 (2001).
\bibitem{28} U. Konopka, G. E. Morfill, and L. Ratke, Phys. Rev. Lett. 84, 891 (2000).
\bibitem{29} S. V. Vladimirov, S. A. Maiorov, and N. F. Cramer, Phys. Rev. E 67, 016407 (2003).
\bibitem{30} M. Lampe, G. Joyce, and G. Ganguli, IEEE Trans. Plasma Sci. 33, 57 (2005).
\bibitem{30a} J.Carstensen, F. Greiner, D. Block, J. Schablinski,W. J. Miloch, and A. Piel, Phys. Plasmas 19, 033702 (2012).
\bibitem{29asb1} V. N. Tsytovich, Y. K. Khodataev, R. Bingham, Comments Plasma Phys. Control. Fusion 17, 249 (1996).
\bibitem{29asb2} S. Hamaguchi, Comments Plasma Phys. Control. Fusion 18, 95 (1997).
\bibitem{31} A. Douglass, V. Land, K. Qiao, L. S. Matthews, and T. Hyde, Phys. Plasmas 19, 013707 (2012). 




\end{thebibliography}
\end{document}